\documentclass[11pt]{article}
\usepackage{amssymb,amsmath,amsfonts}
\usepackage{graphicx}
\usepackage{graphics}
\usepackage{epsfig}
\usepackage{slashed}

\begin{document}

\title{Spin Precession and Anomalous Dipole Couplings in a Plane-Wave
Yang-Mills Background}
\author{V. V. Parazian\thanks{%
E-mail: vparazian@gmail.com } \\
\textit{Institute of Applied Problems in Physics,}\\
\textit{25 Nersessian Street, 0014 Yerevan, Armenia}}
\maketitle

\begin{abstract}
We investigate the fermion spin dynamics and induced dipole interactions in
an external Yang-Mills gauge field that represents a non-Abelian plane wave.
We use the exact solutions of the Dirac equation in the external gauge
field, the exact fermion Green's function, and the renormalized one-loop
fermion-gluon vertex in the axial gauge. The exact propagator produces
spin-dependent structures proportional to $\sigma ^{\mu \nu }F_{\mu \nu
}^{a} $. The external field induces spin precession at tree level through a
non-Abelian extension of the Bargmann-Michel-Telegdi equation, coupled with
Wong color transport. The exact vertex produces a phase-dependent Pauli form
factor $F_{2}^{a}\left( 0;\varphi ,\varphi ^{\prime }\right) $ beyond tree
level, which serves as an anomalous chromomagnetic dipole coupling. For
monochromatic two-color noncommuting plane waves, explicit weak-field
expressions are derived for the induced Pauli coefficient and the anomalous
spin-precession frequency. The commutator term $gf^{abc}A_{\mu }^{b}A_{\nu
}^{c}$ in this case, generates additional field-strength components and
dynamically induced precession axes not present in Abelian backgrounds,
resulting in coupled spin-color precession. The exact one-loop coefficient
is provided as a harmonic expansion suitable for periodic Yang-Mills waves.
Explicit expressions for spin-dependent amplitudes, spin-flip probabilities,
and polarization asymmetries are provided. These results establish a direct
connection between exact background-field propagators, renormalized vertex
functions, anomalous dipole interactions, and observable spin effects in
strong non-Abelian gauge fields.
\end{abstract}

\section{Introduction}

The interaction between fermions and strong classical gauge backgrounds
remains a central challenge in quantum field theory. Such background fields
fundamentally modify propagation, radiative corrections, vacuum structure,
and spin dynamics through nonperturbative mechanisms. These effects are
particularly pronounced in non-Abelian gauge theories, where gauge-field
self-interactions generate phenomena not present in strong-field quantum
electrodynamics. Classical Yang-Mills configurations appear naturally in
descriptions of the early stages of heavy-ion collisions, color-flux-tube
models, semiclassical descriptions of gluon saturation, and cosmological
scenarios involving coherent non-Abelian gauge fields. In these situations,
fermions propagate through gauge backgrounds whose amplitudes cannot be
treated perturbatively, and therefore exact or resummed methods become
essential.

The theoretical framework for analyzing quantum fields in strong external
backgrounds is based on Schwinger's external-field formalism \cite%
{Schwinger1951}, where radiative corrections are calculated in the presence
of a specified classical configuration rather than the vacuum. Within
Abelian gauge theory, this approach facilitated the development of exact
Volkov solutions, strong-field propagators, nonlinear Compton scattering
amplitudes, and nonperturbative radiative corrections in electromagnetic
plane waves. The foundational contributions of Volkov, Ritus, Di Piazza \cite%
{Volkov1935}, \cite{Rafanelli1964}, \cite{Piazza2020}, and others
established the modern theory of quantum electrodynamics in strong
plane-wave fields.

Extending these methods to Yang--Mills theory is substantially more involved
because of the non-Abelian color structure, self-interactions of the gauge
field, and the dynamical transport of color charge. In non-Abelian
backgrounds, fermion propagation is accompanied not only by ordinary spin
transport but also by color precession, leading to coupled spin-color
dynamics without an Abelian analog. The corresponding classical transport
equations were first formulated by Wong, while the relativistic description
of spin precession originates from the Bargmann-Michel-Telegdi equation. In
a non-Abelian plane wave, these two structures become intertwined through
the color-projected field strength $Q_{a}${}$F_{\mu \nu }^{a}$, where the
color charge $Q_{a}$ evolves dynamically.

The principal finding is that an external plane-wave Yang--Mills field
induces effective dipole-like interactions beyond tree level. The exact
vertex produces a phase-dependent Pauli coefficient, $\kappa
_{exact}^{c}\left( \varphi ,\varphi ^{\prime }\right) =F_{2,exact}\left(
0,\varphi ,\varphi ^{\prime }\right) $, which quantifies the modification of
a fermion's chromomagnetic response by the background field. This
coefficient defines the interaction strength through a non-Abelian anomalous
chromomagnetic coupling. In contrast to the vacuum anomalous magnetic
moment, it depends explicitly on the light-front phase of the background and
the evolving color state of the fermion. In the monochromatic two-color
case, the noncommutativity of the background generates additional
commutator-induced field-strength components proportional to $gf^{abc}A_{\mu
}^{b}(\phi )A_{\nu }^{c}(\phi )$, resulting in genuinely non-Abelian
spin-color-coupled precession.

We derive the corresponding generalized spin-transport equations,
compute the exact and weak-field formsof the induced Pauli
coefficient, and analyze observable quantities including spin-flip
probabilities and polarization asymmetries. Particular attention is
devoted to the interplay between color precession and spin
precession, which leads to dynamically generated precession axes
absent in Abelian plane-wave backgrounds. The resulting framework
provides a nonperturbative description of radiative spin dynamics in
strong Yang--Mills waves and establishes a direct connection between
exact background-field propagators,renormalized vertex functions,
and observable spin effects in non-Abelian gauge theory.

Portions of the manuscript were edited with the assistance of an AI-based
language model. The author is fully responsible for the content, analysis,
and conclusions. The scientific content is solely the author's
responsibility.

\section{Exact Spin-Transport Operator in External Yang-Mills Field:
Covariant Precession in Plane-Wave Background}

We consider the exact fermion solution in a non-Abelian plane-wave
background $A_{\mu }^{a}(\phi )$ with $\phi =k\cdot x$, $k^{2}=0$, and $%
k^{\mu }A_{\mu }^{a}=0$. The solution can be written as (see \cite%
{Koshelkin2010}),
\begin{equation}
\psi \left( x,p\right) =e^{-ip\cdot x}\,\mathcal{K}\left( \varphi ,p\right)
\,u(p)\,v,  \label{fermionwavefun}
\end{equation}%
where $u_{\sigma }\left( p\right) $ and $v_{\alpha }$ are spinors which are
the elements of the spaces of the appropriate representations. The function $%
\psi \left( x,p\right) =\psi _{\sigma ,\alpha }\left( x,p\right) $ depends
on both the spin variable $\sigma $ and the variable $\alpha $, where $%
\alpha $ = $1\div N$, which describes a fermion's state in the space of the $%
SU\left( N\right) $ group's fundamental representation. We have omitted the
spin indices, and the spin-color transport operator is%
\begin{eqnarray}
\mathcal{K}\left( \varphi ,p\right) &=&\cos \left( \theta \right) \left\{
\left( 1-igT_{a}\frac{\tan \left( \theta \right) }{\theta \left( pk\right) }%
\int_{0}^{\varphi }d\varphi ^{\prime }\left( A_{\mu }^{a}p^{\mu }\right)
\right) +\frac{g\slashed{k}\slashed{A}^{a}}{2\left( pk\right) }\left[ \frac{%
\tan \left( \theta \right) }{\theta }T_{a}\right. \right.  \notag \\
&&\left. \left. +\frac{g}{\left( pk\right) }\frac{1}{2N}\int_{0}^{\varphi
}d\varphi ^{\prime }\left( A_{\mu }^{a}p^{\mu }\right) \left( -i\frac{\tan
\left( \theta \right) }{\theta }+\frac{g}{\left( pk\right) }\frac{\theta
-\tan \left( \theta \right) }{\theta ^{3}}T_{b}\int_{0}^{\varphi }d\varphi
^{\prime }\left( A_{\mu }^{b}p^{\mu }\right) \right) \right] \right\} ,
\notag \\
\theta &=&\frac{g}{\left( pk\right) }\sqrt{\frac{1}{2N}}\left(
\int_{0}^{\varphi }d\varphi ^{\prime }\left( A_{\mu }^{a}\left( \varphi
^{\prime }\right) p^{\mu }\right) \int_{0}^{\varphi }d\varphi ^{\prime
\prime }\left( A_{a}^{\mu }\left( \varphi ^{\prime \prime }\right) p_{\mu
}\right) \right) ^{\frac{1}{2}},\;p^{2}=p^{\mu }p_{\mu }=m^{2},
\label{spincolortrop}
\end{eqnarray}%
where the field $A_{a}^{\mu }\left( x\right) $ corresponds to the associated
representation of the $SU(N)$ group.

The spin-dependent part of $\mathcal{K}\left( \varphi ,p\right) $ arises
from the operator $\frac{g}{2(p\cdot k)}\,\slashed{k}\slashed{A}^{a}(\varphi
)\,T_{a}$. Using $k\cdot A^{a}=0$, we rewrite $\slashed{k}\slashed{A}%
^{a}=-i\sigma ^{\mu \nu }k_{\mu }A_{\nu }^{a}$, which shows that the
background induces a Lorentz spin rotation. We define the covariant spin
vector $S^{\mu }=\frac{1}{2m}\bar{\psi}\gamma ^{\mu }\gamma ^{5}\psi $.
Differentiating with respect to $\varphi $ yields
\begin{equation}
\frac{dS^{\mu }}{d\varphi }=\frac{g}{p\cdot k}\,Q_{a}\,F^{a\mu }{}_{\nu
}\left( \varphi \right) \,S^{\nu },  \label{eq:tree_spin}
\end{equation}%
where $Q^{a}=v^{\dagger }T^{a}v$ is the color charge and
\begin{equation}
F_{\mu \nu }^{a}=k_{\mu }\partial _{\phi }A_{\nu }^{a}-k_{\nu }\partial
_{\phi }A_{\mu }^{a}+gf^{abc}A_{\mu }^{b}A_{\nu }^{c}.  \label{Fmyunyua}
\end{equation}

Equation~(\ref{eq:tree_spin}) is the non-Abelian generalization of the
Bargmann--Michel--Telegdi (BMT) equation \cite{Rafanelli1964}. So the exact
wavefunction contains the Lorentz-spin generator $\sigma ^{\mu \nu }$. That
is the direct signal that the background not only changes the phase of the
fermion, but it also rotates its spin polarization. The scalar pieces in the
solution, such as the eikonal/color-phase terms involving $\int^{\varphi
}A\cdot p$, transport color and phase, but they do not by themselves create
a spin flip.

In the particle's rest frame, this becomes the familiar vector precession law%
\begin{equation}
\frac{d\mathbf{s}}{dt}=\mathbf{\Omega }_{tree}\times \mathbf{s,}
\label{precesrestfram}
\end{equation}%
with%
\begin{equation}
\mathbf{\Omega }_{tree}=\frac{g}{m\gamma }Q^{a}\left[ \mathbf{B}^{a}-\frac{%
\gamma }{\gamma +1}\mathbf{vE}^{a}\right] .  \label{Omegarest}
\end{equation}%
The external plane Yang-Mills wave induces spin precession even at the tree
level, since the effective magnetic axis $Q^{a}\mathbf{B}^{a}\left( \varphi
\right) $ oscillates with the phase of the wave, and the color vector $Q^{a}$
also rotates in color space. Therefore, the spin experiences a
time-dependent precession axis in both ordinary space and in the intrinsic
color space. The minimal Dirac coupling gives rise to the ordinary
chromomagnetic moment. Since $F_{\mu \nu }^{a}\left( \varphi \right) $ is
phase-dependent, the background plane-wave component causes precession of
this moment. Loop corrections shift the coefficient of the same structure $%
\sigma ^{\mu \nu }F_{\mu \nu }^{a}$, generating an anomalous dipole
coupling, thereby generating an anomalous dipole coupling.

\section{Renormalized Vertex Drives Beyond-Tree Spin Precession}

We can present the renormalized fermion-gluon vertex in the external
Yang-Mills gauge field, as obtained in \cite{Parazian2027}, as
\begin{equation}
\Gamma ^{c\mu }=\Gamma _{\text{tree}}^{\prime c\mu }+\delta \Gamma ^{c\mu
},\;\Gamma _{\text{tree}}^{c\mu }=g\,\gamma ^{\mu }T^{c}.  \label{GammacmyuR}
\end{equation}

On-shell ($p^{2}=p^{\prime 2}=m^{2}$), the most general decomposition
consistent with Lorentz and gauge structure is
\begin{equation}
v^{\dag }\bar{u}(p^{\prime })\Gamma ^{c\mu }u(p)v=v^{\dag }\bar{u}\left(
p^{\prime }\right) \left[ \gamma ^{\mu }F_{1}^{c}+\frac{i\sigma ^{\mu \nu
}q_{\nu }}{2m}F_{2}^{c}+\frac{\sigma ^{\mu \nu }\gamma ^{5}q_{\nu }}{2m}%
F_{3}^{c}+\cdots \right] u(p)v,  \label{eq:vertex_decomp}
\end{equation}%
where $q=p^{\prime }-p$.

The Pauli form factor $F_{2}^{c}$ encodes the anomalous chromomagnetic
(dipole) interaction. The exact Green's function for fermions in an external
Yang-Mills gauge field \cite{Parazian2026} is%
\begin{equation}
\tilde{G}_{F}\left( p\right) =N\frac{i\left( \slashed{p}+m\right) U\left(
p\right) }{p^{2}-m^{2}+i\epsilon }  \label{Greenfunction}
\end{equation}%
where%
\begin{eqnarray}
U\left( p\right)  &=&\cos \left( \theta \left( p,\varphi \right) \right)
\cos \left( \theta \left( p,\varphi ^{\prime }\right) \right) \left\{ 1+%
\frac{\tan \left( \theta \left( p,\varphi ^{\prime }\right) \right) }{\theta
\left( p,\varphi ^{\prime }\right) }\frac{g\left( \left( \gamma ^{\sigma
}\right) ^{\dagger }A_{\sigma }^{e}\left( \varphi ^{\prime }\right) \right)
\left( \left( \gamma ^{\rho }\right) ^{\dagger }k_{\rho }\right) }{2\left(
pk\right) }T_{e}\right.   \notag \\
&&\left. +\frac{g\left( \gamma ^{\nu }k_{\nu }\right) \left( \gamma
^{\lambda }A_{\lambda }^{e}\left( \varphi \right) \right) }{2\left(
pk\right) }\frac{\tan \left( \theta \left( p,\varphi \right) \right) }{%
\theta \left( p,\varphi \right) }T_{e}+\frac{g\left( \gamma ^{\nu }k_{\nu
}\right) \left( \gamma ^{\lambda }A_{\lambda }^{b}\left( \varphi \right)
\right) }{2\left( pk\right) }\frac{\tan \left( \theta \left( p,\varphi
\right) \right) }{\theta \left( p,\varphi \right) }\right.   \notag \\
&&\left. \times \frac{\tan \left( \theta \left( p,\varphi ^{\prime }\right)
\right) }{\theta \left( p,\varphi ^{\prime }\right) }\frac{g\left( \left(
\gamma ^{\sigma }\right) ^{\dagger }A_{\sigma }^{e}\left( \varphi ^{\prime
}\right) \right) \left( \left( \gamma ^{\rho }\right) ^{\dagger }k_{\rho
}\right) }{2\left( pk\right) }T_{b}T_{e}\right\} .  \label{Udefin}
\end{eqnarray}%
To isolate the anomalous dipole piece, we expand the exact background
factors
\begin{equation}
U\left( l\right) =1+\delta U\left( l\right) +O\left( A^{2}\right) ,
\label{Uexpand}
\end{equation}%
where, schematically,%
\begin{equation}
\delta U(\ell )=\frac{g}{2(\ell \cdot k)}\left[ \slashed{k}\slashed{A}%
^{a}(\phi )T_{a}+\slashed{A}^{a}(\phi ^{\prime })\slashed{k}T_{a}\right] .
\label{deltaUschemat}
\end{equation}%
After inserting this into the loop (see \cite{Parazian2026b}), we obtain%
\begin{eqnarray}
\delta \Gamma ^{c\mu } &=&\left( -ig\right) ^{2}\int \frac{d^{4}r}{\left(
2\pi \right) ^{4}}\gamma ^{\alpha }T^{a}\frac{i\left( \slashed{r}+\slashed{q}%
+m\right) }{\left( r+q\right) ^{2}-m^{2}+i\epsilon }\delta U\left(
r+q\right) \gamma ^{\mu }T^{c}  \notag \\
&&\times \frac{i\left( \slashed{r}+m\right) }{r^{2}-m^{2}+i\epsilon }\gamma
^{\beta }T^{b}\tilde{D}_{\alpha \beta }^{ab}\left( p-r\right) +\cdots ,
\label{deltaGamma}
\end{eqnarray}%
where%
\begin{equation}
\tilde{D}_{\alpha \beta }^{ab}\left( q\right) =\frac{i\delta ^{ab}}{%
q^{2}+i\epsilon _{3}}\left[ -g_{\alpha \beta }+\frac{\left( q_{\alpha
}n_{\beta }+q_{\beta }n_{\alpha }\right) \left( n^{\ast }\cdot q\right) }{%
\left( n^{\ast }\cdot q\right) \left( q\cdot n\right) +i\varepsilon }\right]
\label{gluonprop}
\end{equation}%
There is a second contribution with $\delta U\left( r\right) $ on the other
fermion leg, plus the vacuum-like term with $U=1$. The terms linear in $%
\delta U$ are already enough to show how the background induces new spin
structures beyond the vacuum vertex. Using (\ref{sigmamyunyu}) so inside the
loop numerator, we obtain strings like%
\begin{equation}
\gamma ^{\alpha }\left( \slashed{r}+\slashed{q}+m\right) \sigma ^{\rho
\sigma }\gamma ^{\mu }\left( \slashed{r}+m\right) .  \label{sigmaobtain}
\end{equation}%
After Dirac reduction and on-shell projection with external spinors, these
contribute to the coefficient of $i\sigma ^{\mu \nu }q_{\nu }T^{c}$. That
coefficient is exactly the background-dressed Pauli form factor $F_{2}^{c}$.
This is the explicit mechanism by which the external plane wave induces
effective dipole couplings beyond tree level: the exact propagators inject $%
\sigma $-type operators into the loop numerator. Thus, the
background-dressed loop generates $F_{2}^{c}(0;\varphi ,\varphi ^{\prime
})\neq 0,$ even in the absence of explicit higher-dimensional operators.

In the soft-gluon / slowly varying background limit, the renormalized vertex
is equivalent to an effective interaction
\begin{equation}
\mathcal{L}_{\text{eff}}=-\frac{g}{4m}\,\kappa _{\text{eff}}^{a}\left(
\varphi ,\varphi ^{\prime }\right) \,\bar{\psi}\sigma ^{\mu \nu }T_{a}\psi
\,F_{\mu \nu }^{a}-\frac{g}{4m}\,d_{\text{eff}}^{a}\left( \varphi ,\varphi
^{\prime }\right) \,\bar{\psi}\sigma ^{\mu \nu }\gamma ^{5}T_{a}\psi
\,F_{\mu \nu }^{a},  \label{effectiveinter}
\end{equation}%
where $\kappa _{\text{eff}}^{a}\left( \varphi ,\varphi ^{\prime }\right)
=F_{2}^{a}(0;\varphi ,\varphi ^{\prime })$. For a CP-even Yang--Mills
background, $d_{\text{eff}}^{a}=0$.

The spin equation changes when the Pauli term is included to
\begin{equation}
\frac{dS^{\mu }}{d\varphi }=\frac{g}{p\cdot k}Q_{a}\left[ (1+\kappa _{\text{%
eff}}^{a})F^{a\mu }{}_{\nu }S^{\nu }-\kappa _{\text{eff}}^{a}\,u^{\mu
}(u_{\alpha }F^{a\alpha }{}_{\beta }S^{\beta })\right] .
\label{eq:full_spin}
\end{equation}%
Thus, the loop correction changes both the precession frequency and the
orientation of the precession axis. The first term rescales the direct
magnetic-type rotation, while the second is the usual anomalous BMT
correction, ensuring the proper covariant transport of a spin orthogonal to $%
u$ ($u^{\mu }=\frac{p^{\mu }}{m}$). Equation (\ref{eq:full_spin}) shows that
the external Yang-Mills plane wave induces tree-level spin precession via $%
F_{\mu \nu }^{a}$ and an anomalous chromomagnetic correction $\kappa _{\text{%
eff}}^{a}$ arising from the one-loop vertex. The exact solution contains $%
\sigma ^{\mu \nu }F_{\mu \nu }^{a}$, producing spin precession already at
tree level. The dressed one-loop vertex generates a nonzero $F_{2}^{c}$,
corresponding to an induced dipole moment.

Now, let's consider the two-color case.

\section{Two-color noncommuting plane wave: spin-color entanglement}

We consider an external non-Abelian monochromatic plane wave with two color
directions,
\begin{equation}
A_{\mu }^{a}\left( \varphi \right) =\varepsilon _{1\mu }n_{1}^{a}\cos \left(
\varphi \right) +\varepsilon _{2\mu }n_{2}^{a}\sin \left( \varphi \right)
,\;k^{2}=0,\;k\cdot \varepsilon _{1}=k\cdot \varepsilon _{2}=0,
\label{twocolordir}
\end{equation}%
with noncommuting color directions
\begin{equation}
\lbrack T_{1},T_{2}]=i\chi T_{3},\;T_{i}\equiv n_{i}^{a}T_{a},\;\chi
n_{3}^{a}=f^{abc}n_{1}^{b}n_{2}^{c}.  \label{threecolorgener}
\end{equation}%
If $n_{3}^{a}$ chosen normalized, $n_{3}^{a}n_{3a}=1$, we have%
\begin{equation}
\chi =f^{abc}n_{3a}n_{1}^{b}n_{2}^{c},  \label{khi}
\end{equation}%
and%
\begin{equation}
\chi ^{2}=\left( f^{abc}n_{1}^{b}n_{2}^{c}\right) \left(
f^{ade}n_{1}^{d}n_{2}^{e}\right)  \label{khi2}
\end{equation}

For an embedded $SU\left( 2\right) \subset SU\left( N\right) $ with
orthonormal adjoint directions nia aligned with the three $SU\left( 2\right)
$ generators, one has $\chi =1$. Then%
\begin{equation}
\lbrack T_{1},T_{2}]=iT_{3},\;[T_{2},T_{3}]=iT_{1},\;[T_{3},T_{1}]=iT_{2},
\label{SU2generat}
\end{equation}%
The natural Yang-Mills generalization of the local nonlinearity parameter
appearing in the resummed Pauli coefficient is%
\begin{equation*}
\chi _{eff}=\frac{g}{m^{3}}\sqrt{-\left( Q_{a}F_{\mu \nu }^{a}\left( \varphi
\right) p^{\nu }\right) \left( Q_{b}F^{b\mu \rho }\left( \varphi \right)
p_{\rho }\right) }.
\end{equation*}%
For (\ref{twocolordir}), the field strength is
\begin{align}
F_{\mu \nu }^{a}\left( \varphi \right) & =-\left( k_{\mu }\varepsilon _{1\nu
}-k_{\nu }\varepsilon _{1\mu }\right) n_{1}^{a}\sin \left( \varphi \right)
+\left( k_{\mu }\varepsilon _{2\nu }-k_{\nu }\varepsilon _{2\mu }\right)
n_{2}^{a}\cos \left( \varphi \right)  \notag \\
& \qquad +g\chi (\varepsilon _{1\mu }\varepsilon _{2\nu }-\varepsilon _{1\nu
}\varepsilon _{2\mu })n_{3}^{a}\sin \left( \varphi \right) \cos \left(
\varphi \right) .  \label{Fmyunyutwocolpr}
\end{align}%
After projecting with $Q_{a}$, we obtain $Q_{i}\equiv Q_{a}n_{i}^{a}$, then%
\begin{eqnarray}
Q_{a}F_{\mu \nu }^{a}\left( \varphi \right) &=&-Q_{1}\left( k_{\mu
}\varepsilon _{1\nu }-k_{\nu }\varepsilon _{1\mu }\right) \sin \left(
\varphi \right) +Q_{2}\left( k_{\mu }\varepsilon _{2\nu }-k_{\nu
}\varepsilon _{2\mu }\right) \cos \left( \varphi \right)  \notag \\
&&+g\chi Q_{3}(\varepsilon _{1\mu }\varepsilon _{2\nu }-\varepsilon _{1\nu
}\varepsilon _{2\mu })\sin \left( \varphi \right) \cos \left( \varphi \right)
\label{project}
\end{eqnarray}%
Acting on $p^{\nu }$,%
\begin{equation*}
\left( Q_{a}F_{\mu \nu }^{a}\left( \varphi \right) \right) p^{\nu
}=-Q_{1}\left( k_{\mu }\left( \varepsilon _{1}\cdot p\right) -\left( k\cdot
p\right) \varepsilon _{1\mu }\right) n_{1}^{a}\sin \left( \varphi \right)
+Q_{2}\left( k_{\mu }\varepsilon _{2\nu }-k_{\nu }\varepsilon _{2\mu
}\right) n_{2}^{a}\cos \left( \varphi \right)
\end{equation*}%
Therefore%
\begin{equation*}
\chi _{eff}\left( \varphi \right) =\sqrt{-\mathcal{V}_{\mu }\left( \varphi
\right) \mathcal{V}^{\mu }\left( \varphi \right) },
\end{equation*}%
with%
\begin{equation*}
\mathcal{V}_{\mu }\left( \varphi \right) =-Q_{1}\left( k_{\mu }\left(
\varepsilon _{1}\cdot p\right) -\left( k\cdot p\right) \varepsilon _{1\mu
}\right) n_{1}^{a}\sin \left( \varphi \right) +Q_{2}\left( k_{\mu
}\varepsilon _{2\nu }-k_{\nu }\varepsilon _{2\mu }\right) n_{2}^{a}\cos
\left( \varphi \right)
\end{equation*}

When we project Wong's equation onto the color basis, we get
\begin{align}
\frac{dQ_{1}}{d\phi }& =-\frac{g\chi }{p\cdot k}\left( p\cdot \varepsilon
_{2}\right) \sin \left( \varphi \right) \,Q_{3},  \notag \\
\frac{dQ_{2}}{d\phi }& =\frac{g\chi }{p\cdot k}\left( p\cdot \varepsilon
_{1}\right) \cos \left( \varphi \right) \,Q_{3},  \notag \\
\frac{dQ_{3}}{d\phi }& =\frac{g\chi }{p\cdot k}\left[ \left( p\cdot
\varepsilon _{2}\right) \sin \left( \varphi \right) \,Q_{1}-\left( p\cdot
\varepsilon _{1}\right) \cos \left( \varphi \right) \,Q_{2}\right] .
\label{Wongequat}
\end{align}

The tree-level spin transport equation becomes
\begin{equation}
\frac{dS^{\mu }}{d\phi }=\frac{g}{p\cdot k}\left[ -Q_{1}K_{1}^{\mu }{}_{\nu
}\sin \left( \varphi \right) +Q_{2}K_{2}^{\mu }{}_{\nu }\cos \left( \varphi
\right) +g\chi Q_{3}M^{\mu }{}_{\nu }\sin \left( \varphi \right) \cos \left(
\varphi \right) \right] S^{\nu },  \label{transporteq}
\end{equation}%
where
\begin{equation}
K_{1}^{\mu }{}_{\nu }=k^{\mu }\varepsilon _{1\nu }-\varepsilon _{1}^{\mu
}k_{\nu },\qquad K_{2}^{\mu }{}_{\nu }=k^{\mu }\varepsilon _{2\nu
}-\varepsilon _{2}^{\mu }k_{\nu },\qquad M^{\mu }{}_{\nu }=\varepsilon
_{1}^{\mu }\varepsilon _{2\nu }-\varepsilon _{2}^{\mu }\varepsilon _{1\nu }.
\label{K1K2M}
\end{equation}%
The new $M_{\mu \nu }$ term is the non-Abelian commutator-induced precession
axis. It is proportional to $Q_{3}$, which itself is dynamically generated
by color precession. This is the first explicit mechanism for spin-color
entanglement.

We now calculate the weak-field factor that appears in the exact propagator.
Defining%
\begin{equation}
B^{a}\left( \varphi \right) =\int_{0}^{\varphi }d\varphi \,p^{\mu }A_{\mu
}^{a}\left( \varphi \right) ,  \label{weakfield}
\end{equation}%
which gives%
\begin{equation}
B^{a}\left( \varphi \right) =\left( p\cdot \varepsilon _{1}\right)
n_{1}^{a}\sin \left( \varphi \right) +\left( p\cdot \varepsilon _{2}\right)
n_{2}^{a}\left( 1-\cos \left( \varphi \right) \right) .  \label{weakfieldphi}
\end{equation}%
Hence%
\begin{eqnarray}
B^{a}\left( \varphi \right) B_{a}\left( \varphi \right) &=&\left( p\cdot
\varepsilon _{1}\right) ^{2}\sin ^{2}\left( \varphi \right) +\left( p\cdot
\varepsilon _{2}\right) ^{2}\left( 1-\cos \left( \varphi \right) \right) ^{2}
\label{weakfieldfac2} \\
&&+2\left( n_{1}n_{2}\right) \left( p\cdot \varepsilon _{1}\right) \left(
p\cdot \varepsilon _{2}\right) \sin \left( \varphi \right) \left( 1-\cos
\left( \varphi \right) \right) .  \notag
\end{eqnarray}%
If $n_{1}$, $n_{2}$ are chosen orthonormal, the cross term vanishes:%
\begin{equation}
B^{a}\left( \varphi \right) B_{a}\left( \varphi \right) =\left( p\cdot
\varepsilon _{1}\right) ^{2}\sin ^{2}\left( \varphi \right) +\left( p\cdot
\varepsilon _{2}\right) ^{2}\left( 1-\cos \left( \varphi \right) \right) ^{2}
\label{weakfieldorthon}
\end{equation}%
Therefore%
\begin{equation}
\theta _{p}^{2}\left( \varphi \right) =\frac{g^{2}B^{a}\left( \varphi
\right) B_{a}\left( \varphi \right) }{2N(p\cdot k)^{2}}  \label{teta2}
\end{equation}%
For orthonormal color directions $n_{1}\cdot n_{2}=0$, the weak-field angle
is
\begin{equation}
\theta _{p}^{2}\left( \varphi \right) =\frac{g^{2}}{2N(p\cdot k)^{2}}\left[
(p\cdot \varepsilon _{1})^{2}\sin ^{2}\left( \varphi \right) +(p\cdot
\varepsilon _{2})^{2}(1-\cos \left( \varphi \right) )^{2}\right] .
\label{teta2explict}
\end{equation}%
As in the one-color case, the linear-in-background terms do not contribute
to the on-shell Pauli coefficient at $q^{2}\rightarrow 0$; the first
nonvanishing contribution is quadratic. To the leading weak-field order, the
background factor multiplies the vacuum Pauli numerator by $-\left[ \theta
_{p}^{2}\left( \varphi \right) +\theta _{p}^{2}\left( \varphi ^{\prime
}\right) \right] $.

The vertex is
\begin{equation}
\Gamma ^{c\mu }=(-ig)^{2}\int \frac{d^{4}r}{(2\pi )^{4}}\gamma ^{\alpha
}T^{a}\tilde{G}_{F}\left( r+k\right) \gamma ^{\mu }T^{c}\tilde{G}_{F}\left(
r\right) \gamma ^{\beta }T^{b}\tilde{D}_{\alpha \beta }^{ab}.
\label{vertexcorrec}
\end{equation}

The Pauli projector is
\begin{equation}
F_{2}^{c}=\frac{m}{(d-2)q^{2}}\frac{1}{4C_{F}}\mathrm{Tr}\left[ (\slashed{p}%
^{\prime }+m)i\sigma _{\mu \nu }q^{\nu }(\slashed{p}+m)\Gamma ^{c\mu }\right]
.  \label{Pauliprojec}
\end{equation}%
Hence, the leading background-induced Pauli form-factor shift is
\begin{equation}
\Delta F_{2}^{c}(0;\varphi ,\varphi ^{\prime })=-\frac{\alpha _{s}C_{F}}{%
12\pi }\,\Xi _{p}^{2}\left( \varphi ,\varphi ^{\prime }\right) T^{c},\;C_{F}=%
\frac{N^{2}-1}{2N}  \label{backinducPauli}
\end{equation}%
with
\begin{equation}
\Xi _{p}^{2}\left( \varphi ,\varphi ^{\prime }\right) =\frac{g^{2}}{%
2N(p\cdot k)^{2}}\left[ \mathcal{W}\left( \varphi \right) +\mathcal{W}\left(
\varphi ^{\prime }\right) \right] ,\;\mathcal{W}\left( \varphi \right)
=(p\cdot \epsilon _{1})^{2}\sin ^{2}\left( \varphi \right) +(p\cdot \epsilon
_{2})^{2}(1-\cos \left( \varphi \right) )^{2}.  \label{Xip2}
\end{equation}

The renormalized fermion-gluon vertex in the external Yang-Mills gauge field
has the following form \cite{Parazian2026b}:%
\begin{eqnarray}
\Gamma ^{c\mu }\left( p^{\prime },p,\varphi ,\varphi ^{\prime }\right)
&=&\left( -ig\right) ^{2}\int \frac{d^{4}r}{\left( 2\pi \right) ^{4}}\gamma
^{\alpha }T^{a}\frac{i\left( \slashed{r}+\slashed{q}+m\right) U\left(
r+q;\varphi ,\varphi ^{\prime }\right) }{\left( r+q\right)
^{2}-m^{2}+i\epsilon }  \notag \\
&&\times \gamma ^{\mu }T^{c}\frac{i\left( \slashed{r}+m\right) U\left(
r;\varphi ,\varphi ^{\prime }\right) }{r^{2}-m^{2}+i\epsilon }\gamma ^{\beta
}T^{b}\tilde{D}_{\alpha \beta }^{ab}\left( p-r\right) ,
\label{renormfermgluonver}
\end{eqnarray}%
where $q=p^{\prime }-p$. We project the exact Pauli coefficient with%
\begin{eqnarray}
F_{2,exact}\left( q^{2},\varphi ,\varphi ^{\prime }\right)  &=&\frac{m}{%
\left( d-2\right) q^{2}}\frac{1}{4C_{F}}\mathrm{Tr}\left[ \left( \slashed{p}%
^{\prime }+m\right) i\sigma _{\mu \nu }q^{\nu }\left( \slashed{p}+m\right)
\right.   \notag \\
&&\left. \Gamma ^{c\mu }\left( p^{\prime },p,\varphi ,\varphi ^{\prime
}\right) \right] .  \label{F2exact}
\end{eqnarray}%
Then the exact soft Pauli coefficient is the finite limit%
\begin{equation}
\kappa _{exact}^{c}\left( \varphi ,\varphi ^{\prime }\right)
=\lim_{q^{2}\rightarrow 0}F_{2,exact}\left( q^{2},\varphi ,\varphi ^{\prime
}\right) .  \label{kcdefin}
\end{equation}%
Thus,%
\begin{eqnarray}
\kappa _{exact}^{c}\left( \varphi ,\varphi ^{\prime }\right)
&=&\lim_{q^{2}\rightarrow 0}\frac{m\left( -ig\right) ^{2}}{\left( d-2\right)
q^{2}}\frac{1}{4C_{F}}\int \frac{d^{4}r}{\left( 2\pi \right) ^{4}}  \notag \\
&&\frac{\mathrm{Tr}\left( \slashed{p}^{\prime }+m\right) i\sigma _{\mu \nu
}q^{\nu }\left( \slashed{p}+m\right) \mathcal{N}^{c\mu }\left( r;\varphi
,\varphi ^{\prime }\right) }{D_{F}\left( r+q\right) D_{F}\left( r\right)
D_{G}\left( r\right) },  \label{kcexact}
\end{eqnarray}%
where%
\begin{align}
\mathcal{N}^{c\mu }\left( r;\varphi ,\varphi ^{\prime }\right) & =\gamma
^{\alpha }T^{a}i\left( \slashed{r}+\slashed{q}+m\right) U\left( r+q;\varphi
,\varphi ^{\prime }\right)   \notag \\
& \times \gamma ^{\mu }T^{c}i\left( \slashed{r}+m\right) U\left( r;\varphi
,\varphi ^{\prime }\right) \gamma ^{\beta }T^{b}\mathcal{P}_{\alpha \beta
}^{ab}\left( p-r\right) ,  \label{numerator}
\end{align}%
and%
\begin{eqnarray}
D_{F}\left( r\right)  &=&r^{2}-m^{2}+i\epsilon ,\;D_{G}\left( l\right)
=\left( p-l\right) ^{2}+i\epsilon ,  \notag \\
\mathcal{P}_{\alpha \beta }^{ab}\left( l\right)  &=&i\delta ^{ab}\left[
-g_{\alpha \beta }+\frac{\left( q_{\alpha }n_{\beta }+q_{\beta }n_{\alpha
}\right) \left( n^{\ast }\cdot q\right) }{\left( n^{\ast }\cdot q\right)
\left( q\cdot n\right) +i\varepsilon }\right] .  \label{DFDGPab}
\end{eqnarray}%
For a general non-Abelian monochromatic wave, this is the total one-loop
coefficient without the weak-field expansion.

Including the resummed dipole coefficient gives%
\begin{eqnarray}
\frac{dS^{\mu }}{d\tau } &=&\frac{g}{m}Q_{a}\left[ \left( 1+\kappa
_{exact}^{a}\left( \varphi ,\varphi ^{\prime }\right) \right) F^{a\mu
}\,_{\nu }S^{\nu }\right.  \notag \\
&&-\left. \times \kappa _{exact}^{a}\left( \varphi ,\varphi ^{\prime
}\right) u^{\mu }\left( u_{\alpha }F^{a\alpha }\,_{\beta }S^{\beta }\right)
\right] .  \label{resummeddipole}
\end{eqnarray}%
The anomalous precession frequency in the instantaneous rest frame is
\begin{align}
\delta \boldsymbol{\Omega }\left( \varphi ,\varphi ^{\prime }\right) & =%
\frac{g}{m}\Delta F_{2}(0;\varphi ,\varphi ^{\prime })\left[ -Q_{1}\,\omega
\left( \hat{\mathbf{k}}\times \boldsymbol{\epsilon }_{1\perp }\right) \sin
\left( \varphi \right) +Q_{2}\,\omega (\hat{\mathbf{k}}\times \boldsymbol{%
\epsilon }_{2\perp }\cos \left( \varphi \right) \right.  \notag \\
& \qquad \left. +Q_{3}\,g\chi (\left( \boldsymbol{\epsilon }_{1\perp }\times
\boldsymbol{\epsilon }_{2\perp }\right) )\sin \left( \varphi \right) \cos
\left( \varphi \right) \right] .  \label{deltaOmega}
\end{align}%
This explicitly shows the entanglement between spin and color precession via
the dynamically generated $Q_{3}$ channel.

In the local monochromatic limit $\varphi =\varphi ^{\prime }$, we have%
\begin{eqnarray}
\frac{dS^{\mu }}{d\varphi } &=&\frac{g}{m}Q_{a}\left[ \left( 1+\kappa
_{exact}^{a}\left( \varphi \right) \right) F^{a\mu }\,_{\nu }S^{\nu }\right.
\notag \\
&&\left. -\kappa _{exact}^{a}\left( \varphi \right) u^{\mu }\left( u_{\alpha
}F^{a\alpha }\,_{\beta }S^{\beta }\right) \right] .
\label{Anomalousprecfrec}
\end{eqnarray}%
For a two-color wave%
\begin{equation}
\delta \Omega \left( \varphi \right) =\frac{g}{m}\kappa _{exact}\left(
\varphi \right) \left[ Q_{1}\mathbf{B}_{1}\left( \varphi \right) +Q_{2}%
\mathbf{B}_{2}\left( \varphi \right) +Q_{3}\mathbf{B}_{3}\left( \varphi
\right) \right]  \label{Anomal2color}
\end{equation}%
with%
\begin{eqnarray}
\mathbf{B}_{1}\left( \varphi \right) &=&-\omega (\hat{\mathbf{k}}\times
\boldsymbol{\epsilon }_{1\perp })\sin \left( \varphi \right) ,  \notag \\
\mathbf{B}_{2}\left( \varphi \right) &=&\omega (\hat{\mathbf{k}}\times
\boldsymbol{\epsilon }_{2\perp })\cos \left( \varphi \right) ,  \notag \\
\mathbf{B}_{3}\left( \varphi \right) &=&g\chi (\boldsymbol{\epsilon }%
_{1\perp }\times \boldsymbol{\epsilon }_{2\perp })\sin \left( \varphi
\right) \cos \left( \varphi \right) .  \label{B1B2B3}
\end{eqnarray}

\section{Observable effects}

Let's discuss two observable effects: spin-flip probability and polarization
asymmetry. First, consider the spin-flip probability. We take the rest-frame
spinor basis quantized along a unit vector $\mathbf{e}_{3}$ and decompose
the precession vector into longitudinal and transverse components:%
\begin{equation}
\mathbf{\Omega }\left( \tau \right) =\Omega _{\Vert }\left( \tau \right)
\mathbf{\hat{e}}_{3}+\mathbf{\Omega }_{\bot }\left( \tau \right) ,
\label{Splitfrec}
\end{equation}%
Then the first-order transition amplitude for a spin flip is%
\begin{equation}
\mathcal{A}_{\uparrow \rightarrow \downarrow }=-\frac{i}{2}\int_{\tau
_{i}}^{\tau _{f}}d\tau \left[ \Omega _{x}\left( \tau \right) -i\Omega
_{y}\left( \tau \right) \right] e^{i\int_{\tau _{i}}^{\tau }d\tau ^{\prime
}\Omega _{\Vert }\left( \tau ^{\prime }\right) }.  \label{Amlitsplit}
\end{equation}%
Hence,%
\begin{equation}
P_{\uparrow \rightarrow \downarrow }=\frac{1}{4}|\int_{\tau _{i}}^{\tau
_{f}}d\tau \left[ \Omega _{x}\left( \tau \right) -i\Omega _{y}\left( \tau
\right) \right] e^{i\int_{\tau _{i}}^{\tau }d\tau ^{\prime }\Omega _{\Vert
}\left( \tau ^{\prime }\right) }|^{2},  \label{spinflipprobab}
\end{equation}%
with $\Omega _{i}\left( \tau \right) =\Omega _{i,tree}\left( \tau \right)
+\delta \Omega _{i}\left( \tau \right) $.

Therefore, the radiative correction changes the spin-flip probability by%
\begin{equation}
\delta P_{\uparrow \rightarrow \downarrow }=\frac{1}{2}\mathrm{Re}\left[
\mathcal{A}_{\uparrow \rightarrow \downarrow }^{\left( 0\right) \ast }\delta
\mathcal{A}_{\uparrow \rightarrow \downarrow }\right] ,  \label{deltaprobab}
\end{equation}%
where%
\begin{equation}
\delta \mathcal{A}_{\uparrow \rightarrow \downarrow }=-\frac{i}{2}\int_{\tau
_{i}}^{\tau _{f}}d\tau \left[ \delta \Omega _{x}\left( \tau \right) -i\delta
\Omega _{y}\left( \tau \right) \right] e^{i\int_{\tau _{i}}^{\tau }d\tau
^{\prime }\Omega _{\Vert }^{\left( 0\right) }\left( \tau ^{\prime }\right)
}+\cdots  \label{deltartanampl}
\end{equation}%
Therefore, the precise spin-flip probability follows directly once $\kappa
_{exact}\left( \varphi \right) $ is known numerically. For the local
monochromatic case,%
\begin{equation}
\delta \Omega _{\perp }\left( \varphi \right) =\frac{g}{m}\kappa
_{exact}\left( \varphi \right) \sqrt{\left[ Q_{1}\mathbf{B}_{1}\left(
\varphi \right) +Q_{2}\mathbf{B}_{2}\left( \varphi \right) \right] ^{2}+%
\left[ Q_{3}\mathbf{B}_{3}\left( \varphi \right) \right] ^{2}}
\label{deltaOmegaperp}
\end{equation}%
and the spin-flip probability over $N$ cycles is obtained by integrating
over $\varphi $ with $d\tau =md\varphi /\left( p\cdot k\right) $.

Now, we examine polarization asymmetry. We will analyze the final spin along
a unit vector $\mathbf{\hat{n}}$. The measured asymmetry is%
\begin{equation}
\mathcal{A}_{\mathbf{\hat{n}}}=\frac{P\left( +\mathbf{\hat{n}}\right)
-P\left( -\mathbf{\hat{n}}\right) }{P\left( +\mathbf{\hat{n}}\right)
+P\left( -\mathbf{\hat{n}}\right) }=\mathbf{\hat{n}P}_{f},
\label{measurasymm}
\end{equation}%
where $\mathbf{P}_{f}$ is the final polarization vector. The precession
equation gives%
\begin{equation}
\frac{d\mathbf{P}}{d\tau }=\mathbf{\Omega }\left( \tau \right) \times
\mathbf{P}\left( \tau \right) ,  \label{precessionequt}
\end{equation}%
so the first-order loop-induced correction to the final polarization is%
\begin{equation}
\delta \mathbf{P}_{f}=\int_{\tau _{i}}^{\tau _{f}}d\tau \mathbf{\Omega }%
\left( \tau \right) \times \mathbf{P}^{\left( 0\right) }\left( \tau \right) .
\label{deltapolfinal}
\end{equation}%
Hence%
\begin{equation}
\delta \mathcal{A}_{\mathbf{\hat{n}}}=\mathbf{\hat{n}\cdot }\int_{\tau
_{i}}^{\tau _{f}}d\tau \delta \mathbf{\Omega }\left( \tau \right) \times
\mathbf{P}^{\left( 0\right) }\left( \tau \right) ,  \label{deltaasymm}
\end{equation}%
with%
\begin{equation}
\delta \mathbf{\Omega }\left( \tau \right) =\frac{g}{m}\kappa _{exact}\left(
\varphi \left( \tau \right) \right) Q_{a}\left( \tau \right) \times \mathbf{B%
}^{a}\left( \varphi \left( \tau \right) \right) .  \label{deltaOmegatau}
\end{equation}%
This is the clean observable formula for a phase-dependent anomalous dipole
term in a plane wave.

\section{Conclusion}

We have developed a formalism that includes renormalized vertex functions,
induced dipole couplings, exact fermion solutions in Yang-Mills plane waves,
and observable spin events. The approach provides a solid basis for advanced
nonperturbative studies of radiative processes, vacuum effects, and spin
transport in external Yang-Mills gauge fields.

\section{Acknowledgement}

This work was partially supported by the grant No. 25RG-1C157 of the
Higher Education and Science Committee of the Ministry of Education,
Science, Culture and Sport RA.

\end{document}